\documentclass[10pt,showpacs,nofootinbib,aps,superscriptaddress,eqsecnum,prd,notitlepage,showkeys]{revtex4-1}
 \usepackage[utf8]{inputenc}
\usepackage{amsmath}
\usepackage{amssymb}
\usepackage{graphicx}
\usepackage{esint}
\usepackage{textcomp}

\begin{document}
 
\title{Retrieving the scattering strength using noise correlation in a reverberant medium}

\author{Aida Hejazi Nooghabi}

\affiliation{Sorbonne Universit\'e, CNRS-INSU, Institut des Sciences de la Terre Paris, ISTeP UMR 7193, F-75005 Paris, France}

\affiliation{ESPCI Paris, PSL Research University, Institut Langevin, 1 rue Jussieu, F-75005, Paris, France}

\author{Lapo Boschi}

\affiliation{Sorbonne Universit\'e, CNRS-INSU, Institut des Sciences de la Terre Paris, ISTeP UMR 7193, F-75005 Paris, France}

\author{Philippe Roux}

\affiliation{Laboratoire ISTERRE, Universit\'e Grenoble Alpes, CNRS, 1380 rue de la Piscine, F-38000, Grenoble, France}

\author{Julien de Rosny}

\affiliation{ESPCI Paris, PSL Research University, Institut Langevin, 1 rue Jussieu,
F-75005, Paris, France}

\begin{abstract}
We explore how ambient-noise interferometry can serve to track variations in the distribution of scatterers within a medium. Experiments are conducted in the presence and absence of a 
single scatterer that can occupy different positions within the propagation medium. We calculate the similarity between the Green's functions obtained in these 
cases (\textit{i.e.}, with \textit{versus} without the scatterer) as a measure of the temporal variations in the Green's function, which shows exponential decay with respect to time. We 
also develop a formula that relates this decay to the scattering strength of the scatterer. The formulae are experimentally validated using experiments in a strongly reverberating duralumin 
plate. This proposed method for finding the scattering strength of the scatterer is based on information carried by the coda of the multiple echo arrivals. We confirm the reliability of 
this approach through a conventional method for the measurement of the scattering strength that is based on direct arrivals.
\end{abstract}

\maketitle
\section{Introduction}
In complex media, waves are scattered or reflected many times before their extinction, producing random-like and time-dispersed wave-fields. However, the late parts of the echo arrivals remain 
strongly imprinted upon by the propagation medium. Long-range and infinite-range correlations \cite{Akkermans} and back-scattering enhancement are several of the expressions of this property. With a 
complex medium that changes with time, the dynamics of the fluctuation patterns of the late arrivals are directly related to the medium dynamics. For instance, coda-wave interferometry
\cite{snieder2006} focuses on the coda evolution when a large-scale modification of a complex medium occurs, such as changes in velocity associated with changes in temperature. The coda is the part 
of the transient response (\textit{i.e.}, Green's function) that results from multiple scattering interferences. To locate a weak and local modification in a multiple-scattering medium, the LOCADIFF algorithm 
was proposed by \cite{locadiff2}, which is based on diffusion theory \cite{sheng}. LOCADIFF involves an inversion procedure that consists of the spatial sensitivity map of the wave-field with respect to local 
changes in the medium. Diffusive-wave spectroscopy \cite{pine1988} is another method based on analysis of light-intensity fluctuations 
when all of the scatterers are moving randomly. This idea was originally developed in the context of electromagnetic waves, but has since been applied to acoustics
\cite{daws2000}, \cite{daws2002}. Contrary to diffusive-wave spectroscopy, which works in the frequency domain, diffusive acoustic-wave spectroscopy is based on temporal analysis of the late time fluctuations 
of the coda. More recently, the coda generated in a reverberant cavity when one or several scatterers are moving was analyzed \cite{derosny2001}, \cite{rosny2003}. This method, known as diffuse reverberant 
acoustic-wave spectroscopy (DRAWS), has been shown to be promising for finding scattering cross-sections \cite{derosny2001} and the displacement 
magnitudes \cite{conti2005} of a discrete set of moving scatterers. The scattering cross-section is a property of a scatterer that quantifies its strength averaged over all of the incident angles.
% and represents the wave field intensity contribution of a 
% scatterer when it intercepts the incident wave.

All of these techniques, from coda-wave interferometry to DRAWS, are based on the measurement of the two-point Green's functions between a set of sources and receivers, either in the frequency domain or in the time domain. However, when dealing with time-domain techniques, it might not be possible to have access to point-like and transient sources to directly measure the Green's function. Through the advent of noise interferometry, the Green's function between two points in a medium can be retrieved by cross-correlation of the noise recorded at these locations; \textit{i.e.}, without the need for an active source. The first, pioneering, validation of this theoretical result in the experimental domain was carried out by \cite{Weaver2001}, who propagated elastic waves through a block of aluminum, and explained their observations by the principle of equipartition of energy over different modes. Later, this method found applications in different domains, such as ocean acoustics (\textit{e.g.}, \cite{Roux2004}), seismology (\textit{e.g.},  \cite{Campillo2014,Boschi2015}), and medical imaging (\textit{e.g.}, \cite{Sabra2007}). 
A number of studies have shown that this method can be used to estimate the coda in a multiple-scattering medium or in a reverberant cavity (\cite{larose2008}, \cite{julienepl}, \cite{Hej2017}). Following these studies, passive coda-wave interferometry was successfully applied in seismology, based on the analysis of coda estimated from noise correlations (\textit{e.g.}, \cite{Brenguier2008}, \cite{Brenguier2014}). In reverberating media, several studies have shown that it is possible to passively localize a scatterer on a thin elastic plate from the analysis of the direct path recovered by noise correlation (\textit{e.g.}, \cite{Chehami-localization}). However, the sensitivity of the method is poor, and does not yet provide the cross-section of the scatterer. 

The present study is based on the DRAWS approach, and we show that the scattering cross-section can be estimated in reverberant media without controlled sources. 
This study demonstrates the reliability of DRAWS based on passive retrieval of the Green's function for monitoring of the variations in the medium, with possible applications to acoustic or seismic waves. We carry out experiments on a thin 
duralumin plate and recover the Green's function through cross-correlation.    
In the first test, we repeat our experiment with and without a single scatterer. In the second test, we repeat the experiment before and after changing the position of the scatterer. In both cases, the decorrelation in the coda of these passively recovered Green's functions is tracked by computing the similarity coefficient between the Green's functions. We next develop a formalism that is similar to that proposed by \cite{rosny2003}, and the scattering cross-section is estimated. The possibility to track the temporal fluctuations in a coda caused by the appearance of a scatterer, and not only by a change in its position, is an extension, first attempted here, of classical DRAWS. Moreover, DRAWS is applied to highly dispersive plate waves for the first time here. 

The passive estimation of the scattering cross-section is also double checked: 
first by the use of active DRAWS, and then by classical estimation of the cross-section based on the measurement of the scattering phase function from direct plane-wave illumination. There is a close match between the passive measurements and the two active measurements. 

This paper is structured as follows: section II explains in detail the theoretical approach, and section III presents the experimental set-up that is used for the passive/ active scattering cross-section estimations. In section IV, the experimental results are presented and discussed, before the final conclusion in section V.

\section{Theoretical analysis}
In contrast to coda-wave interferometry, where nonlocal distributed variations in wave speed are of interest, in the present study, we focus on the effects of very sharp and localized heterogeneities, which we refer to as scatterers.

In this section, we discuss two slightly different cases for estimation of the scattering cross-section of a scatterer based on the coda of the source-to-receiver Green's 
function. As demonstrated by \cite{rosny2003}, the basic assumption in this formalism is that the scattering mean free path ($l_s$) is much larger compared to the dimension of the cavity. First, we study the case where the Green's function of the medium in the presence of the scatterer is compared to that obtained in its absence. This will be referred to as case I in the following.

In a strongly reverberating medium, the time-domain Green's function (\textit{i.e.}, impulse response) between a source at $x_{1}$ and a receiver at $x_{2}$ ($G(x_1,x_2,t)$) can be described as:
\begin{equation}\label{Gi}
G_{w/,i}(x_1,x_2,t) = G_{0i}(x_1,x_2,t)\exp(-t/2\tau) + s_{i}(t)
\end{equation}
where $i$ is an integer index that indicates a certain source-receiver within a database. The subscript $w/$ serves to remind us that one scatterer is present somewhere in the medium. $G_{0i}(x_1,x_2,t)$ is the impulse response in the absence of the scatterer, which is known to introduce an exponential decay with a decay time of $\tau$ \cite{rosny2003}. $s_{i}(t)$ is a term due to the presence of the scatterer \cite{rosny2003}. 

In the second set of experiments, the scatterer is removed. For such a situation, the exponential decay due to the presence of the scatterer will disappear, and hence $G_{w/o,i}(x_1,x_2,t)$ can be written as:
\begin{equation}\label{Gsans}
G_{w/o,i}(x_1,x_2,t) = G_{0i}(x_1,x_2,t),
\end{equation}
where the subscript $w/o$ indicates 'without scatterer'. 
We are interested in measuring the similarity between the Green's functions given by Equations (\ref{Gi}) and (\ref{Gsans}). We repeat the measurements of $G_{w/}$ and $G_{w/o}$, and take their mean ($<>$) over all of the available source-receiver pairs, to find the time-dependent Pearson correlation coefficient (hereafter referred to as 'similarity'), 
\begin{equation}
 S(t) =\frac{< G_{w/,i}(x_1,x_2,t)G_{w/o,i}(x_1,x_2,t)>}{\sqrt{<(G_{w/,i}(x_1,x_2,t))^{2}>}\sqrt{<(G_{w/o,i}(x_1,x_2,t))^{2}>}}\label{pearson}
\end{equation}
Substituting Equations (\ref{Gi}) and (\ref{Gsans}) into Equation (\ref{pearson}) and defining the residual terms as uncorrelated, it turns out that
\begin{equation}
S(t) = \exp \left(-t/2\tau\right)\label{swwo}.
\end{equation}
In a diluted scattering medium where the density of scatterers is not high, the scattering mean free path ($l_{s}$) that is defined as the mean distance between two scatterers can be expressed as \cite{sheng}:
\begin{equation}
 l_{s} = \frac{1}{n\sigma}\label{ls}
\end{equation}
where $n$ is the density of scatterers and $\sigma$ is the scattering cross-section of the scatterer. The scattering mean free path is one of the parameters that describes the propagation of waves in a multiple-scattering medium (\textit{e.g.}, \cite{Derode1995}). In the present case, we deal with an unusual scattering regime, because as in the DRAWS technique, an important number of reverberations occur between two scattering events. Following that, the decay time is found by dividing Equation \ref{ls} by the wave speed $c$,
\begin{equation}
 \tau = \frac{1}{cn\sigma}.
\end{equation}
Equation (\ref{swwo}) can be rewritten as:
\begin{equation}\label{exp}
 S(t) = \exp\left(-tcn\sigma/2\right).
\end{equation}
As these experiments are performed on a duralumin plate, the energy propagates in the form of Lamb waves, which are highly dispersive; \textit{i.e.}, the wave speed changes with the frequency. In the low-frequency regime, the phase speed ($V_\phi$) of the fundamental mode of antisymmetric Lamb waves at a given angular frequency ($\omega$) and for a plate of thickness $e$, density $\rho$, and flexural rigidity $D$ is given by \cite{Royer2000}
\begin{equation}\label{phase_speed}
 V_\phi(\omega) = \sqrt{\omega}\left(\frac{D}{\rho e}\right)^\frac{1}{4}.
\end{equation}
In the measurements here, we are interested in the velocity of the propagation of wave packets and not of a single phase. So, $c$ in Equation (\ref{exp}) should be replaced by the group velocity ($V_g(\omega)$), which in the low-frequency regime is twice the phase speed given by Equation (\ref{phase_speed}). Substituting for $c$ leads to
\begin{equation}\label{swwo2}
 S(t) = \exp \left(-\left(\frac{D}{\rho e}\right)^\frac{1}{4}t\sqrt{\omega_0}n\sigma\right)
\end{equation}
where $\omega_0$ is the mean angular frequency in the bandwidth of interest. 

For case II, the Green's functions are compared when the position of the scatterer has changed between successive acquisitions. To avoid repetition, we only describe here the differences with respect to case I. Here, as the scatterer is always present in the medium, the Green's function between the same source and receiver in the second experiment is
\begin{equation}\label{Gj}
G_{w/,i'}(x_1,x_2,t) = G_{0i'}(x_1,x_2,t)\exp(-t/2\tau) + s_{i'}(t).
\end{equation}
Similar to case I, $i$ indicates the source-receiver pair and the change in the position of the scatterer is indicated by the prime sign.

Replacing $G_{w/o,i}(x_1,x_2,t)$ with $G_{w/,i'}(x_1,x_2,t)$ in Equation (\ref{pearson}), and following the same procedure as above, the similarity coefficient is written as:
\begin{equation}\label{sf}
 S(t) = \exp \left(-2\left(\frac{D}{\rho e}\right)^\frac{1}{4}t\sqrt{\omega_0}n\sigma\right).
\end{equation}
Equations (\ref{swwo2}) and (\ref{sf}) suggest that the scattering cross-section can be determined based on experimental/ field observations of $S$. In other words, if a single or a set of unknown scatterers are added to the plate, the measures of the Green's function carried out before and after the introduction of the scatterers can recover the strength of the scatterers. This is also true where the Green's functions 
are measured following the displacement of the scatterer. In the following, we experimentally verify this theoretical analysis, and we confirm its validity through
comparison with a conventional method.

\section{Experimental validation}
\begin{figure}[h!]
\begin{center}
  \includegraphics[trim=3cm  6cm  6cm  5cm,clip=true,keepaspectratio,scale=0.8]{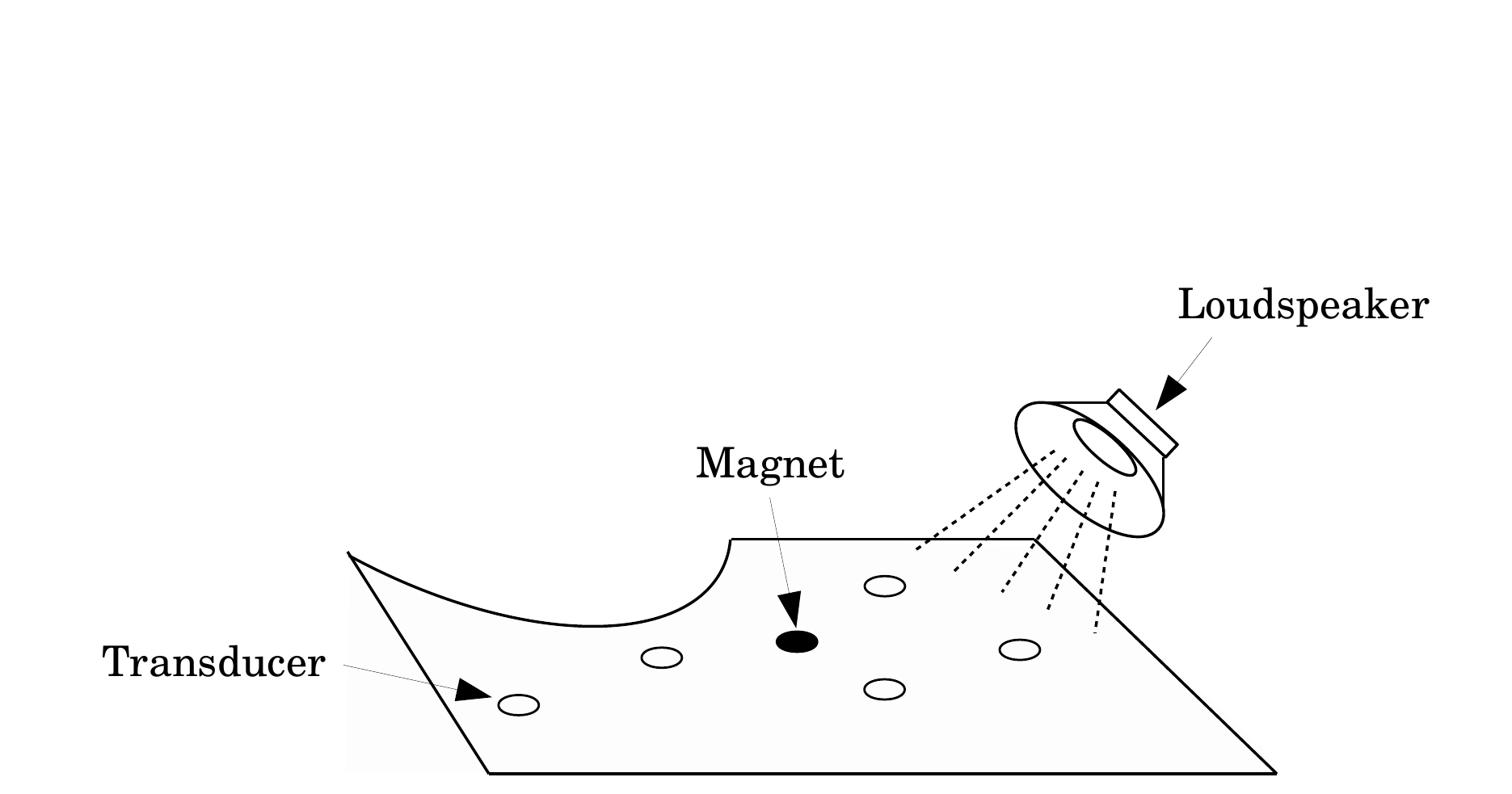}
 \caption{Sinai billiard-shaped plate used in the experimental set-up to find the scattering cross-section by interferometry. The plate is suspended horizontally and transducers are connected to an audio card.}\label{fig1}
 \end{center}
\end{figure}

To validate the proposed formulae for cases I and II, acoustic experiments are carried out on plates. The experimental set-up is as shown in Fig. \ref{fig1}, and consists of a quarter Sinai billiard \cite{Sinai1970} shaped duralumin plate of 75 $\times$ 75 $\times$ 0.3 cm. The plate is suspended using two thin vertical supports to provide free boundary conditions. There are five piezoelectric transducers attached to the plate surface. A pair of cylindrical magnets of 12 mm diameter and 5 mm height are attached to the plate at the exact same location on both sides of the plate. This pair of magnets acts as a single scatterer that is symmetrical with respect to the mid-plane of the plate, and redirects the propagating Lamb waves without mode conversion. 

To retrieve the Green's function between each pair of transducers, ambient-noise interferometry is used, which is valid provided that the propagating waves
travel along all directions with equal probability. This condition is what is referred to as a 'diffuse wave-field' (see \cite{Kinsler1999}, section 12.1; see also \cite{Boschi2015}, \cite{Campillo2014} for reviews of seismic applications of noise interferometry). 
In a reverberating or strongly scattering medium, whatever the source, the coda tends to be at least approximately diffuse after a sufficient number of reverberations and/or scattering events have occurred. In this set-up, the geometry of the plate is selected so as to eliminate any preferential direction of the propagation. This has been shown to result in the condition of diffusivity being more easily met \cite{Mortessagne}. 

\begin{figure}[h!]
 \includegraphics[keepaspectratio,scale=0.4]{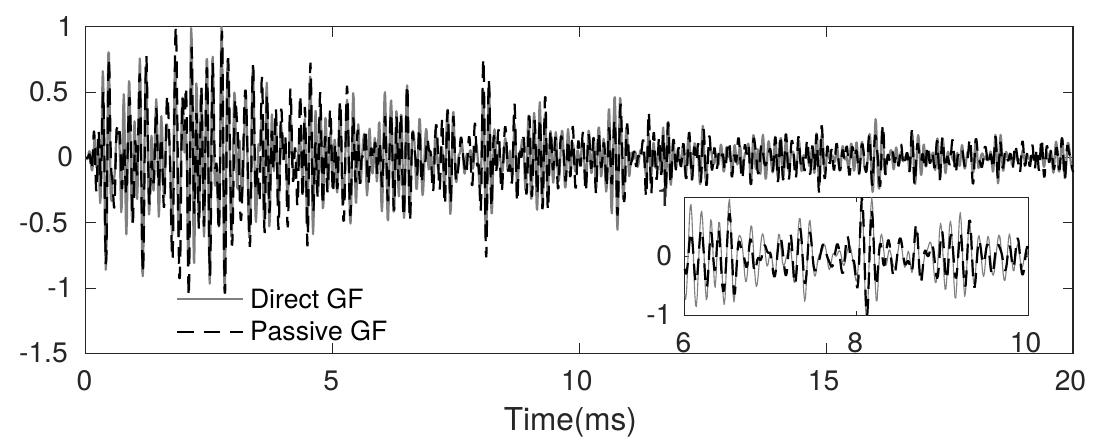}
 \caption{Comparison between actively (gray solid line) and passively (black dashed line) retrieved Green's functions bandpass filtered from 5 kHz to 10 kHz and normalized by maximum amplitude. The inset shows the zoom from 6 ms to 10 ms that corresponds to $\sim$4 m of propagation in the reverberating plate.}\label{comp}
\end{figure}

Pseudo-random noise is emitted by a loudspeaker (Ryght). The loudspeaker is moved erratically at 10 cm above the surface during the acquisition. At the same time, the propagating waves are recorded at all of the deployed transducers. The Green's function between each pair of transducers is determined by cross-correlation of the corresponding pair of long-duration recordings. The retrieved Green's function includes the direct and coda waves that propagate between the transducers, which coincides 
with the signal that one transducer would record if the other were an impulsive source. The Green's function obtained by this method will be referred to here as the 'passive' Green's function, as there is no active source involved. The advantage of using transducers attached to the plate is that they can be used both as sources and receivers. This allows the direct measurement of the Green's function between each pair of transducers. The Green's functions so obtained will be referred to here as 'direct' or 'active', because in this case each transducer has the role of an active source. We use the active Green's functions as reference to determine the reliability of the Green's functions obtained by cross-correlation. The source signal is a linearly varying frequency chirp between 1500 Hz and 90000 Hz. An example of the comparison between active and passive Green's functions is shown in Figure \ref{comp} in the frequency band of 5 kHz to 10 kHz.
Fig. \ref{comp} shows a close match between these two Green's functions. Once the high-quality Green's functions are retrieved passively, we apply them to Equation \ref{pearson} to find the similarity $S(t)$, and then Equations \ref{swwo2} or \ref{sf} are implemented to determine the scattering cross-section. The corresponding procedure to find the scattering cross-section after the retrieval of the Green's function is explained in the following.

\subsection{Reconstruction of scattering cross-section by interferometry when a scatterer appears in the propagation medium (case I)}

\begin{figure}[h!]
\begin{center}
  \includegraphics[trim=1cm  0.0cm  0cm  20cm,clip=true,keepaspectratio,scale=0.5]{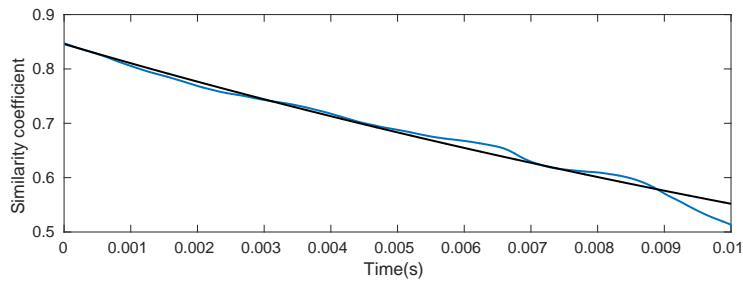}
 \caption{Similarity coefficient between the 'passive' Green's functions 'with' and 'without' the magnets, filtered between 7.5 kHz and 12.5 kHz (blue), and the fitting based on Equation \ref{swwo2} (black).}\label{decay}
 \end{center}
\end{figure}

Given a discrete set of five transducers, we deal with Green's function between 10 combinations of two receivers. Following the passive estimation of the Green's functions in the presence of the magnets, we next remove the magnets and repeat the same measurement. For each of 10 pairs of transducers, we calculate the similarity coefficient between the Green's function in the presence \textit{versus} absence of the scatterer. We then take the mean of the similarity coefficients over all of the 10 possibilities according to Equation \ref{pearson}. These measurements are repeated for different magnet positions, and show that the similarity coefficients in the presence \textit{versus} absence of the scatterer are independent of the position of the magnet. To take into account the dispersion effect with the plate waves, the similarity coefficient is calculated for Green's functions that are filtered over a limited frequency band of 5 kHz, with the central frequency varying from 5 kHz to 17.5 kHz.  The group velocity for the central frequency in each bandwidth is substituted in the calculation, and $\sigma(\omega)$ is determined by fitting the exponential decay according to Equation \ref{swwo2}. An example of the decay of the similarity coefficient with time for one pair of transducers at 10 kHz is shown in Figure \ref{decay}, along with the fit according to Equation \ref{swwo2}.
All of the experiments that are performed for the interferometry method are then repeated with active signals, to determine the scattering cross-section based on these 
actively recovered Green's functions. A comparison of the values of the scattering cross-sections based on the active and passive Green's functions in case I is shown in Figure \ref{d12}.

\subsection{Reconstruction of the scattering cross-section by interferometry for displacement of a scatterer (case II)}

As opposed to case I, where the variations in the Green's function were tracked in the presence and absence of the magnets, here the scatterer is always present in the medium and the similarity coefficient is calculated when the position of the scatterer (magnets) has changed from one acquisition to the other.
%The change in the position of the magnet is equivalent to the movement of the fish in the study of \cite{derosny2001}.
We follow an experimental measurement approach similar to that described in the previous section for case I. The mean of the similarity coefficient over different pairs of transducers is calculated in the same way, over the different frequency bands. The exponential decay according to Equation \ref{sf} is then fitted, and finally, the scattering cross-section is obtained as a function of the frequency. The $\sigma(\omega)$ for the actively measured Green's functions are also measured (see Fig. \ref{d12}). In the following subsection, the scattering cross-section of the magnet is measured with an entirely different approach that is based on the direct arrivals.

\subsection{Reference measurement of the scattering cross-section}
\begin{figure}[h!]
\centering
 \includegraphics[trim=3cm  6cm  5cm  1cm,clip=true,keepaspectratio,scale=0.6]{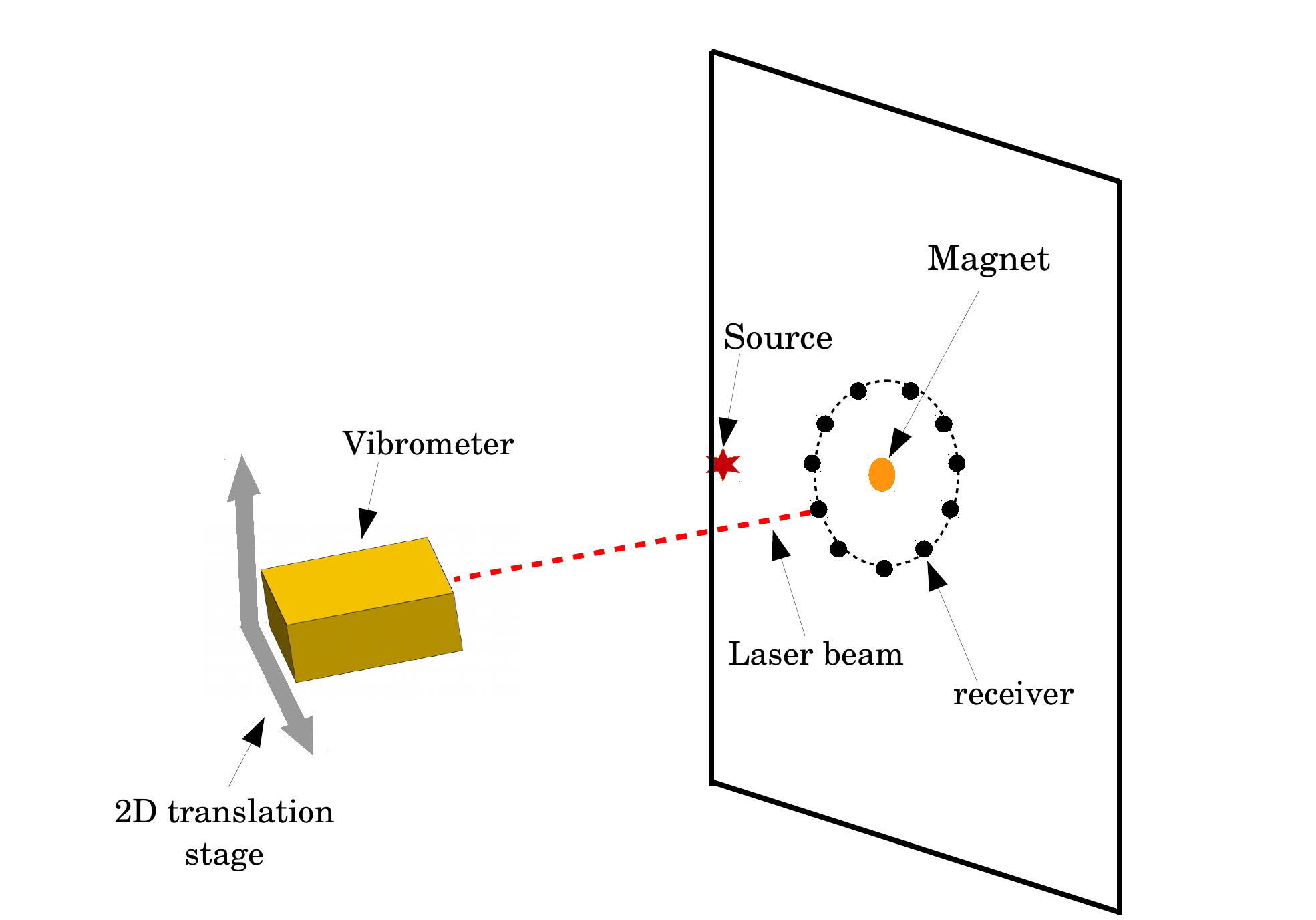}
 \caption{Experimental set-up described in section III.C. The scattering cross-section is extracted from the time-domain analysis of the direct arrivals.}\label{large_plate}
\end{figure}

\begin{figure}[h!]
\centering
 \includegraphics[trim=0cm  6.8cm  0.0cm  5.5cm,clip=true,keepaspectratio,scale=0.65]{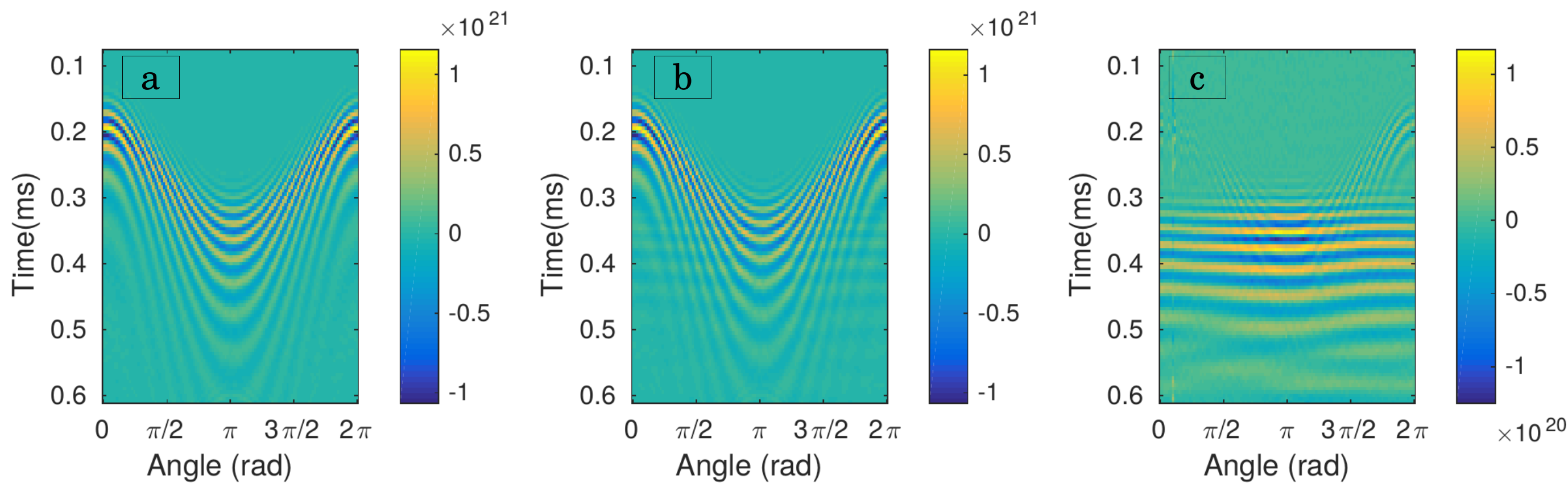}
 \caption{ (a) Recorded direct wave-field on a homogeneous plate (\textit{i.e.}, plate with no magnets). (b) Recorded direct wave-field in the presence of the magnets, with  12 mm diameter and 5 mm height. (c) Subtraction of the recorded wave-fields in the presence and absence of the magnets.}\label{sad}
\end{figure}

For comparison, a conventional method is also used to measure the scattering cross-section of a cylindrical steel magnet. This method relies on the measurement of the scattering wave-field contribution associated to the direct waves. This experiment is carried out on a larger plate of dimensions 1.5 $\times$ 1.0 $\times$ 0.003 m (see Fig. \ref{large_plate}). The use of the larger plate is to ensure an approximately plane incident wave. The source is a piezoelectric transducer that emits the same 
source signal as before. The magnets (\textit{i.e.,} the scatterer) are placed at the middle of the plate, on each side of it, and the velocity field is scanned with a laser interferometer for a circle of radius of $R = 15$ cm, where the center is the center of the magnet (see Fig. \ref{sad}b for the recorded wave-field). Next, the magnets are removed and the same scanning procedure is carried out (Fig. \ref{sad}a).
 Subtracting the part of the recorded velocity field that contains only the direct-arrival information for the presence ($W_{w/}$) and absence ($W_{w/o}$) of the magnets, we have access to the scattered wave-field as a function of the angle \cite{Norris1995}.
The scattering cross-section is deduced from the scattering phase function ($f$) as a measure of the angular distribution of the scattering wave amplitude as a function of frequency, and is defined as the following in the frequency domain \cite{sheng}:
\begin{equation}
\begin{aligned}
f(\theta,\omega) = \frac{W_{w/}(\theta,\omega)-W_{w/o}(\theta,\omega)}{|W_{inc}(\omega)|}\sqrt{R}
\end{aligned}
\end{equation}
where $R$ is the radius of the circle of the scan and $W_{inc}$ is the incident wave-field. From $f(\theta,\omega)$, we obtain the total scattering cross-section $\sigma$ according to \cite{sheng}:
\begin{equation}
\sigma = \int_{0}^{2\pi}|f(\theta,\omega)|^2 d\theta .
\end{equation}
The $\sigma(\omega)$ values obtained by this method for the scattering cross-section serve as the reference to evaluate the $\sigma(\omega)$ values previously 
extracted from the coda of the reverberated signals, as described above for cases I and II. In the following section, the results obtained with the different set-ups and methodologies are compared. 

\section{Results and discussion}
\begin{figure}[h!]
\centering
 \includegraphics[trim=1cm 0.5cm 0cm 21cm,clip=true,keepaspectratio,scale=0.7]{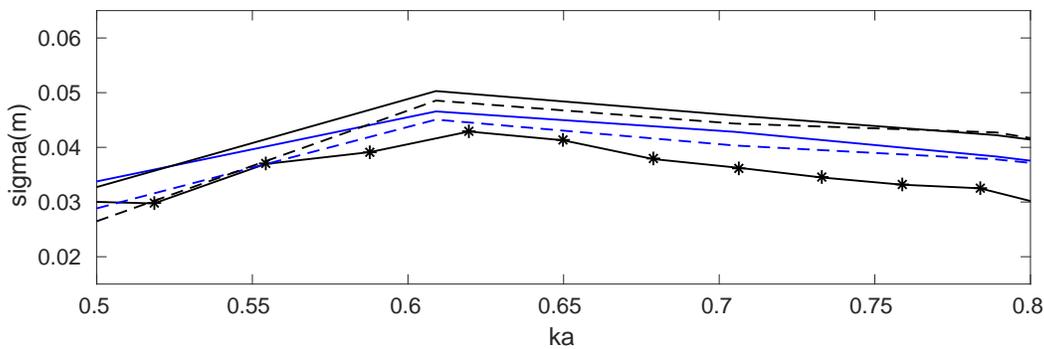}
 \caption{Total scattering cross-sections obtained for the cylindrical magnet of 12 mm diameter and 5 mm height for the different methods and different data. Solid line with stars, reference curve obtained by the conventional method based on direct arrivals; 
 solid blue line, case I for the active Green's function; dashed blue line, case II for the active Green's function; solid black line, case I for the passive Green's function; dashed black line, case II for 
 the passive Green's function.}\label{d12}
\end{figure}

Figure \ref{d12} shows the scattering cross-sections $\sigma(\omega)$ for all of the set-ups and methods here, \textit{versus} the normalized parameter $ka$, where $k$ is the wavenumber and $a$ is the radius of the magnet. The values for both cases I and II, and for both the actively and passively retrieved Green's functions, are 
very close to each other.

In general, the proposed formulae for cases I and II give slightly higher values for $\sigma(\omega)$. We explain this on the basis that these formulae result from an ensemble average, and 
that these experiments are limited to a relatively small number of receiver-receiver pairs (\textit{i.e.}, five receivers, providing 10 pairs). Also, the strong dispersion in the plate and the 
finite size of the plate result in the mixing of the direct and reverberated wavefronts. This complicates the separation of the direct waves in the reference measurement of the scattering cross-section. 
The dispersion is compensated for by applying a filter based on the source-receiver distance. However, this compensation does not work correctly for all frequency bands. Moreover, in the measurements 
based on the coda of the Green's functions, as for the group velocity, we substitute the group velocity of the central frequency in a bandwidth. These approximations give rise to a small bias between 
the different methods of measuring the scattering cross-section. Another observation is that the passive and active Green's functions do not provide similar values, with a small average bias of 
about 0.3 percent. We speculate that this is mostly because the cross-correlation of the diffuse wave-field has still not fully converged to the Green's function.
% This imperfect reconstruction of the Green's function might be due to $S_0$ contribution of excited waves in the active recordings.
This imperfect reconstruction of the Green's function might be due to the contribution of the fundamental symmetric mode of excited waves to the active recordings. When transducers act as receivers, they
are mainly sensitive to the displacements of the antisymmetric mode.
%The emission with a transducer can excite symmetric modes as well as antisymmetric ones, while transducers used as a receiver record the antisymmetric part of the emission. 
The ratio of the antisymmetric to symmetric excited modes decreases with the frequency, which leads to larger bias between the data obtained from the active and passive Green's functions at higher 
frequencies. Limitations such as using a loudspeaker as a noise source prevent the extension of these investigations to higher frequencies, at this stage.

\section{Conclusion}

We have studied here how the scattering cross-section in a reverberating and dispersive medium can be estimated by tracking the small changes in the late coda of the passively retrieved 
Green's function. From a seismology point of view, for instance, we are beginning to evaluate whether, and to what extent, ambient-noise cross-correlation can be used as a monitoring 
tool (\textit{e.g.}, for monitoring volcanoes, earthquakes, landslides). The scatterer is introduced in the medium so that the local elastic parameters such as Young's modulus, density 
and thickness are different from those of the background medium. Ambient signal cross-correlation (also known as noise interferometry) is effective even in the absence of active sources 
or earthquakes, and it only requires the deployment of a set of passive receivers. We analyzed the passively retrieved Green's functions for two different scenarios. In the first, the Green's 
functions obtained in the absence \textit{versus} presence of a single scatterer were compared; in the second case, the Green's functions were compared before and after displacement of a single 
scatterer. These findings were validated by demonstrating that, in both cases, the values for the scattering cross-sections that are based on coda arrivals are similar to those obtained by a 
conventional method based on direct arrivals. 
% The advantage of our proposed method over the conventional method is that it requires less information about the propagating wave field.
% One distinctive point of this study is that here we deal with dispersive waves, and in contrast to previous studies where the scatterer was always present 
% in the medium, we have a case where there is no scatterer in
% the medium. These cases and their corresponding formulae based on the coda arrivals can be used to infer the scattering cross-section of a scatterer 
% provided that the speed of the waves and the density of the scatterers in the medium are known. 

\bibliographystyle{apalike}

\begin{thebibliography}{10}
\bibitem[Akkermans and Montambaux, 2004]{Akkermans}
Akkermans, E. and Montambaux, G. (2004).
\newblock Mesoscopic physics of photons.
\newblock {\em J. Opt. Soc. Am. B}, 21(1):101--112.
\bibitem[Boschi and Weemstra, 2015]{Boschi2015}
Boschi, L. and Weemstra, C. (2015).
\newblock Stationary-phase integrals in the cross correlation of ambient noise.
\newblock {\em Reviews of Geophysics}, 53.
\bibitem[Brenguier et~al., 2014]{Brenguier2014}
Brenguier, F., Campillo, M., Takeda, T., Aoki, Y., Shapiro, N.~M., Briand, X.,
  Emoto, K., and Miyake, H. (2014).
\newblock Mapping pressurized volcanic fluids from induced crustal seismic
  velocity drops.
\newblock {\em Science}, 345(6192):80--82.
\bibitem[Brenguier et~al., 2008]{Brenguier2008}
Brenguier, F., Shapiro, N., Campillo, M., Ferrazzini, V., Duputel, Z., Coutant,
  O., and Nercessian, A. (2008).
\newblock Towards forecasting volcanic eruptions using seismic noise.
\newblock {\em Nature Geoscience}, 1.
\bibitem[Campillo et~al., 2014]{Campillo2014}
Campillo, M., Roux, P., Romanowicz, B., and Dziewonski, A. (2014).
\newblock Seismic imaging and monitoring with ambient noise correlations.
\newblock {\em Treatise on Geophysics}, pages 256--271.
\bibitem[Chehami et~al., 2014]{Chehami-localization}
Chehami, L., Moulin, E., de~Rosny, J., Prada, C., Bou~Matar, O., Benmeddour,
  F., and Assaad, J. (2014).
\newblock Detection and localization of a defect in a reverberant plate using
  acoustic field correlation.
\newblock {\em Journal of Applied Physics}, 115(10):104901.
\bibitem[Conti et~al., 2006]{conti2005}
Conti, S.~G., de~Rosny, J., Roux, P., and Demer, D. (2006).
\newblock Characterization of scatterer motion in a reverberant medium.
\newblock 119:769--776.
\bibitem[Cowan et~al., 2002]{daws2002}
Cowan, M.~L., Jones, I.~P., Page, J.~H., and Weitz, D.~A. (2002).
\newblock Diffusing acoustic wave spectroscopy.
\newblock {\em Phys. Rev. E}, 65:066605.
\bibitem[Cowan et~al., 2000]{daws2000}
Cowan, M.~L., Page, J.~H., and Weitz, D.~A. (2000).
\newblock Velocity fluctuations in fluidized suspensions probed by ultrasonic
  correlation spectroscopy.
\newblock {\em Phys. Rev. Lett.}, 85:453--456.
\bibitem[de~Rosny and Davy, 2014]{julienepl}
de~Rosny, J. and Davy, M. (2014).
\newblock Green's function retrieval and fluctuations of cross density of
  states in multiple-scattering media.
\newblock {\em EPL}, 106(5):54004.
\bibitem[de~Rosny and Roux, 2001]{derosny2001}
de~Rosny, J. and Roux, P. (2001).
\newblock Multiple scattering in a refelcting cavity: Application to fish
  counting in a tank.
\newblock {\em J. Acoust. Soc. Am.}
\bibitem[de~Rosny et~al., 2003]{rosny2003}
de~Rosny, J., Roux, P., Fink, M., and Page, J. (2003).
\newblock Field fluctuation spectroscopy in a reverberant cavity with moving
  scatterers.
\newblock {\em Physical review letters}, 90 9:094302.
\bibitem[Derode et~al., 1995]{Derode1995}
Derode, A., Roux, P., and Fink, M. (1995).
\newblock Robust acoustic time reversal with high-order multiple scattering.
\newblock {\em Physical Review Letters}, 75.
\bibitem[Hejazi~Nooghabi et~al., 2017]{Hej2017}
Hejazi~Nooghabi, A., Boschi, L., Roux, P., and de~Rosny, J. (2017).
\newblock Coda reconstruction from cross-correlation of a diffuse field on thin
  elastic plates.
\newblock {\em Phys. Rev. E}, 96.
\bibitem[Kinsler et~al., 1999]{Kinsler1999}
Kinsler, L., Frey, A., Coppens, A., and Sanders, J. (1999).
\newblock {\em Fundamentals of Acoustics}.
\newblock Wiley, Hoboken, N.J.
\bibitem[Larose et~al., 2008]{larose2008}
Larose, E., Roux, P., Campillo, M., and Derode, A. (2008).
\newblock Fluctuations of correlations and green’s function reconstruction:
  Role of scattering.
\newblock {\em Journal of Applied Physics}, 103(11):114907.
\bibitem[Mortessagne et~al., 1993]{Mortessagne}
Mortessagne, F., Legrand, O., and Sornette, D. (1993).
\newblock Transient chaos in room acoustics.
\newblock {\em Chaos: An Interdisciplinary Journal of Nonlinear Science},
  3(4):529--541.
\bibitem[Norris and Vemula, 1995]{Norris1995}
Norris, A.~N. and Vemula, C. (1995).
\newblock Scattering of flexural waves on thin plates.
\newblock {\em Journal of Sound and Vibration}, 181.
\bibitem[Pine et~al., 1988]{pine1988}
Pine, D.~J., Weitz, D.~A., Chaikin, P.~M., and Herbolzheimer, E. (1988).
\newblock Diffusing wave spectroscopy.
\newblock {\em Phys. Rev. Lett.}, 60:1134--1137.
\bibitem[Rossetto et~al., 2011]{locadiff2}
Rossetto, V., Margerin, L., Plan{\`e}s, T., and Larose, E. (2011).
\newblock {Locating a weak change using diffuse waves: Theoretical approach and
  inversion procedure}.
\newblock {\em {Journal of Applied Physics}}, 109(3):034903.
\bibitem[Roux et~al., 2004]{Roux2004}
Roux, P., Kuperman, W.~A., and Grp, N. (2004).
\newblock Extracting coherent wave fronts from acoustic ambient noise in the
  ocean.
\newblock {\em J. Acoust. Soc. Am.}, 116(4, Part 1):1995--2003.
\bibitem[Royer and Dieulesaint, 2000]{Royer2000}
Royer, D. and Dieulesaint, E. (2000).
\newblock {\em Elastic waves in solids I.}
\newblock Springer-Verlag Berlin, Heidelberg.
\bibitem[Sabra et~al., 2007]{Sabra2007}
Sabra, K., Conti, S., Roux, P., and Kuperman, W. (2007).
\newblock Passive in vivo elastography from skeletal muscle noise.
\newblock {\em Appl. Phys. Lett.}, 90.
\bibitem[Sheng, 1995]{sheng}
Sheng, P. (1995).
\newblock {\em Introduction to Wave Scattering, Localization, and Mesoscopic
  Phenomena}.
\newblock Academic Press, San Diego.
\bibitem[Sinai, 1970]{Sinai1970}
Sinai, Y.~G. (1970).
\newblock Dynamical systems with elastic reflections.
\newblock {\em Russian Mathematical Surveys}, 25:137--189.
\bibitem[Snieder, 2006]{snieder2006}
Snieder, R. (2006).
\newblock The theory of coda wave interferometry.
\newblock {\em Pure and Applied Geophysics}, 163.
\bibitem[Weaver and Lobkis, 2001]{Weaver2001}
Weaver, R. and Lobkis, O.~I. (2001).
\newblock Ultrasonics without a source: Thermal fluctuation correlations at mhz
  frequencies.
\newblock {\em Physical Review Letters}, 87(13).
\end{thebibliography}

\end{document}